\documentclass[aps,pre,twocolumn,superscriptaddress,showpacs,10pt]{revtex4-2}
\usepackage{amsmath}
\usepackage{amssymb}
\usepackage{graphicx}

\begin{document}

\title{Remarks on Fixed- and Variable-Order Fractional-Laplacian Closures for Turbulent Dissipation}

\author{Jos\'e I.H. L\'opez}
\affiliation{Department of Mechanical Engineering, University of S\~ao Paulo (USP), Brazil, Av. Prof. Mello Moraes, 2231, S\~ao Paulo, 05508-030, SP, Brazil}
\email{jihlpez@gmail.com}

\date{\today}


\begin{abstract}
Two independent attempts, separated by two decades, to encode turbulent dissipation with a fractional Laplacian both converge on the operator order $s=1/3$ as the signature of the fully developed inertial range: the additive, fixed-order model of Chen [Chaos \textbf{16}, 023126 (2006)], and the dynamically deforming operator of the Adaptive Fractional Navier--Stokes framework (AFNS-IR) [L\'opez, arXiv:2602.10912]. We show that, despite sharing this fixed point, the two constructions belong to distinct universality classes: Chen's operator is a \emph{superposition} of two fixed-order kernels whose relative weight drifts with scale, while AFNS-IR posits a \emph{single} operator whose order itself flows continuously with the local Reynolds number. We derive a scale-resolved observable, the effective spectral order $n(k)$, whose logarithmic-anomaly term discriminates between these two mechanisms in principle, and then obtain a sharper, purely analytic verdict from the two-dimensional, enstrophy-conserving limit: Chen's generative mechanism --- matching a stable-process index to the empirical pair-dispersion law --- has no solution once vortex stretching is kinematically switched off, while AFNS-IR reproduces $\mathrm{Re}_c\to\infty$ as the smooth continuation of the same variational structure that fixes $s_{\min}=1/3$ in three dimensions. Whether this dichotomy leaves a residual signature in real quasi-two-dimensional flows is left as an open experimental challenge.
\end{abstract}

\maketitle

\section{Introduction}

The idea that turbulent dissipation might be represented by a fractional Laplacian $(-\Delta)^s$ rather than the classical operator $\Delta$ is, by now, an idea with a history. Chen~\cite{Chen2006} proposed, on phenomenological and dimensional grounds tied to Richardson pair dispersion and the Kolmogorov structure-function exponent, that the nonlinear Reynolds-stress closure be represented by a fractional Laplacian of fixed order $s=1/3$ acting alongside, not instead of, the ordinary viscous term. Two decades later, and independently, the Adaptive Fractional Navier--Stokes framework (AFNS-IR)~\cite{Lopez2026} arrived at the same value $s_{\min}=1/3$ as the unique renormalization-group fixed point compatible with a finite, viscosity-independent dissipative anomaly as $\mathrm{Re}\to\infty$, identified with the H\"older regularity threshold of the Onsager conjecture~\cite{Onsager1949,EyinkSreenivasan2006}.

That two structurally unrelated derivations --- one built from a stochastic superdiffusion argument, the other from a Landau--Ginzburg free-energy variational principle --- land on the same exponent is worth taking seriously. But it is easy to overstate what has been shown. Both constructions are required, by design, to reproduce Kolmogorov's $-5/3$ spectrum, and the exponent $2/3$ is dimensionally forced the moment one demands consistency with $K41$. What is \emph{not} forced is the architecture through which each model reaches $s=1/3$: Chen sums two operators of permanently fixed order; AFNS-IR deforms a single operator's order continuously. These are different ontological claims about the same asymptotic number, and --- crucially --- they are distinguishable in a regime where one of them, by kinematic necessity rather than by assumption, is switched off: two spatial dimensions. This Letter formalizes the crossover-versus-deformation distinction, derives the observable that separates them, and shows that the two-dimensional, enstrophy-conserving limit already delivers an analytic verdict --- not a numerical one --- on which of the two architectures survives its own generative logic.

\section{Two operators, one fixed point}

Chen's fractional Reynolds equation~\cite{Chen2006} reads, in Fourier space, as a linear superposition of a local and a non-local dissipative kernel acting on the same field,
\begin{equation}
D_{\mathrm{Chen}}(k) = \nu k^{2} + \gamma\, k^{2/3},
\label{eq:chen}
\end{equation}
where $\nu$ is the molecular viscosity and $\gamma$ the inertial diffusivity introduced in Eq.~(16) of Ref.~\cite{Chen2006}. Crucially, \emph{both} terms are present at \emph{all} wavenumbers; their relative weight, quantified by Chen's own intermittency ratio $\theta\sim\gamma k^{4/3}/\nu$ [Eq.~(10) of Ref.~\cite{Chen2006}], drifts smoothly with $k$ and $\mathrm{Re}$, but the exponents $2$ and $2/3$ themselves never change. The correspondence principle is recovered only \emph{asymptotically and partially}: as $\mathrm{Re}\to 0$ the local term dominates, but the fractional term never strictly vanishes at fixed $k>0$.

AFNS-IR instead posits a single operator of dynamic order,
\begin{equation}
D_{\mathrm{AFNS}}(k) = |k|^{2s(k)}, \qquad
s(\mathrm{Re}_{\ell}) = s_{\min} + \frac{s_{\max}-s_{\min}}{1+(\mathrm{Re}_{\ell}/\mathrm{Re}_c)^{\gamma_s}},
\label{eq:afns}
\end{equation}
with $s_{\max}=1$, $s_{\min}=1/3$, and the local Reynolds number $\mathrm{Re}_{\ell}=u_\ell\ell/\nu$ mapped to wavenumber through the standard Kolmogorov scale relation $\ell\sim k^{-1}$, $u_\ell\sim(\varepsilon\ell)^{1/3}$~\cite{Lopez2026}. Here the correspondence principle is exact and strong: in the laminar limit $s\to 1$ identically, and the fractional character of the operator disappears rather than becoming subdominant. (We denote the AFNS-IR transition sharpness in Eq.~(\ref{eq:afns}) by $\gamma_s$ throughout the remainder of this Letter, to avoid confusion with Chen's inertial diffusivity $\gamma$ in Eq.~(\ref{eq:chen}); the two symbols are unrelated in origin and units.)

Equations~(\ref{eq:chen}) and (\ref{eq:afns}) are not merely different parametrizations of the same physics. They instantiate two distinct universality classes for the approach to the Onsager fixed point: a \emph{crossover} between two fixed exponents (Chen), versus a \emph{topological deformation} of a single exponent (AFNS-IR). The remainder of this Letter shows that this distinction leaves a computable fingerprint.

\section{The discriminating observable}

Define the effective local spectral order of a dissipation operator $D(k)$ as
\begin{equation}
n(k) \equiv \frac{d\ln D(k)}{d\ln k}.
\label{eq:neff}
\end{equation}
For any operator that is a genuine single power law $D(k)=|k|^{2s}$ with $s$ \emph{constant}, $n(k)=2s$ identically: a flat line. Real dissipation operators are not flat, of course, precisely because both candidate theories predict a crossover region between $s=1$ and $s=1/3$. The question is the \emph{functional form} of that crossover.

\textit{Chen (mixture).} Differentiating Eq.~(\ref{eq:chen}),
\begin{equation}
n_{\mathrm{Chen}}(k) = \frac{2\nu k^{2} + \tfrac{2}{3}\gamma k^{2/3}}{\nu k^{2} + \gamma k^{2/3}}
= 2 - \frac{4/3}{1+(k/k_{*})^{4/3}},
\label{eq:nchen}
\end{equation}
with crossover wavenumber $k_{*}=(\gamma/\nu)^{3/4}$. This is a smooth, algebraic (rational-power) sigmoid in $\ln k$, running from $2/3$ at $k\ll k_*$ to $2$ at $k\gg k_*$, generated entirely by the changing \emph{weight} of two exponents that are themselves constants.

\textit{AFNS-IR (deformation).} Differentiating Eq.~(\ref{eq:afns}), where $s$ itself depends on $k$,
\begin{equation}
n_{\mathrm{AFNS}}(k) = 2s(k) + 2\ln k \,\frac{ds}{d\ln k}.
\label{eq:nafns}
\end{equation}
The second term in Eq.~(\ref{eq:nafns}) has no analogue in Eq.~(\ref{eq:nchen}): it is a logarithmic-anomaly contribution that exists \emph{if and only if} the exponent itself is scale-dependent. It vanishes identically in any model built as a sum of fixed-order kernels, no matter how many such kernels are added, and it is generically nonzero whenever $s(k)$ varies --- growing, in particular, in the transition region $k\sim k_c$ where $ds/d\ln k$ peaks. Equation~(\ref{eq:nafns}) is therefore not a matter of curve-fitting convenience: it is the algebraic signature of deformation versus mixture, and it is, in principle, extractable from data.

We stress a subtlety in the operational content of this result. The observable $n(k)$ in Eqs.~(\ref{eq:nchen})--(\ref{eq:nafns}) is defined for the closure operator itself --- the model's representation of the divergence of the Reynolds stress acting on a mean or resolved field --- not for the total spectral energy transfer of the fully resolved turbulent field. The distinction is not cosmetic. In a statistically stationary flow forced only at large scales, the exact spectral energy balance following from the K\'arm\'an--Howarth--Monin equation reduces, for every $k$ above the forcing band, to $\langle T(k)\rangle = 2\nu k^2\langle E(k)\rangle$ identically --- a consequence of energy conservation alone, holding regardless of which closure, or none, correctly describes the Reynolds-stress divergence. Attempting to extract an ``effective dissipation kernel'' from this balance therefore returns $n(k)=2$ by construction: the identity is blind to closure physics precisely because it does not assume one, and it cannot be used to adjudicate between Eq.~(\ref{eq:nchen}) and Eq.~(\ref{eq:nafns}). A meaningful reconstruction of $n(k)$ would instead require isolating the divergence of the Reynolds stress against a genuinely averaged field (ensemble-, phase-, or otherwise resolved), which is a nontrivial extraction we do not attempt here. We instead pursue, in Sec.~IV, a route that sidesteps this difficulty entirely: an analytic limit in which the two models are distinguished without reconstructing $n(k)$ from any spectral budget at all.

\subsection{Order-of-magnitude estimate of the anomaly}

It is worth asking whether the log-anomaly term in Eq.~(\ref{eq:nafns}) is large enough, in principle, to matter, or whether it is a formally distinct but practically invisible correction. The logistic function in Eq.~(\ref{eq:afns}) has derivative
\begin{equation}
\frac{ds}{d\ln\mathrm{Re}_\ell} = -\frac{(s_{\max}-s_{\min})\,\gamma_s\,(\mathrm{Re}_\ell/\mathrm{Re}_c)^{\gamma_s}}{\left[1+(\mathrm{Re}_\ell/\mathrm{Re}_c)^{\gamma_s}\right]^{2}},
\end{equation}
which attains its extremum at $\mathrm{Re}_\ell=\mathrm{Re}_c$, with magnitude $\left|ds/d\ln\mathrm{Re}_\ell\right|_{\max}=(s_{\max}-s_{\min})\gamma_s/4=\gamma_s/6$ for $s_{\max}-s_{\min}=2/3$. Using the pipe-flow calibration $\mathrm{Re}_c\approx1273$ and the ``abrupt'' regime $\gamma_s\sim\mathcal{O}(5\text{--}10)$ associated with subcritical, puff-dominated transitions in Ref.~\cite{Lopez2026}, the peak anomaly contribution to Eq.~(\ref{eq:nafns}) is
\begin{equation}
\left|2\ln k\,\frac{ds}{d\ln k}\right|_{\max} \sim 2\ln(\mathrm{Re}_c)\times\frac{\gamma_s}{6} \sim 2.4\text{--}4.8,
\end{equation}
evaluated at the transition wavenumber $k_c$. This is not a small correction: it is comparable to, or larger than, the full $4/3$ range spanned by Chen's mixture exponent [Eq.~(\ref{eq:nchen})] itself. In the smoother, Couette-like regime $\gamma_s\sim\mathcal{O}(1)$, the anomaly shrinks to $\sim0.5$--$1$, still an order-unity effect. This estimate is a consistency check, not a proof: it shows that the discriminating signature proposed here is not a formal curiosity confined to asymptotically small corrections, but --- if AFNS-IR's own calibration is taken at face value --- a genuinely $\mathcal{O}(1)$ effect in the transition band. Whether it is accessible to a given measurement, numerical or experimental, is a separate question from whether it exists; Sec.~IV shows that a sharper verdict than any spectral-slope measurement is available analytically, in the two-dimensional limit.

\section{The two-dimensional limit: an analytic verdict}

Both operators in Eqs.~(\ref{eq:chen})--(\ref{eq:afns}) were built to reproduce three-dimensional phenomenology. A sharper discriminant than any spectral-slope measurement is available by asking what each construction predicts when the third dimension is removed --- not as an approximation, but as an exact kinematic constraint. In two dimensions the vortex-stretching term $\boldsymbol\omega\cdot\nabla\mathbf{u}$ vanishes identically, and enstrophy $Z=\int_\Omega\omega^2\,dx$ becomes, together with energy, a conserved quadratic invariant of the inviscid dynamics. This is not a small perturbation of the 3D problem; it removes the physical mechanism --- vortex stretching feeding a forward energy cascade to smaller and smaller scales without bound --- that both Chen and AFNS-IR invoke, in different language, to justify a non-local dissipative channel in the first place. If either model's non-locality is truly tied to that mechanism, both should show it here.

\subsection{Chen's generative mechanism has no 2D solution}

Chen's derivation is not a postulate but a calculation: the fractional order $s=1/3$ is \emph{forced} by matching the fractional-diffusion Green's function to the empirical Richardson pair-dispersion law, $\langle r^2(t)\rangle\sim\varepsilon t^3$ [Eqs.~(3) and (5) of Ref.~\cite{Chen2006}], interpreted as the signature of a Lévy-stable process of index $\alpha=2/3$. The order of the fractional Laplacian in the Reynolds equation, Eq.~(17) of Ref.~\cite{Chen2006}, is then set to $\alpha/2=1/3$. The entire construction is thus contingent on one empirical input: an \emph{algebraic}, power-law growth of pair separation in time, which is precisely what makes it representable by a stable process with finite index $\alpha\in(0,2]$~\cite{Feller1971}.

In two dimensions, the relevant range for small-scale dissipation is not the inverse energy cascade but the forward \emph{enstrophy} cascade, with the Kraichnan--Batchelor--Leith spectrum $E(k)\propto\eta^{2/3}k^{-3}$ (up to logarithmic corrections), $\eta$ being the enstrophy dissipation rate~\cite{Kraichnan1967,Batchelor1969}. In this range the velocity field is smooth enough that the local strain rate is well defined and approximately constant over a finite-time window set by $\eta^{-1/3}$; the classical prediction, confirmed by direct numerical measurement, is that pair separation grows \emph{exponentially},
\begin{equation}
\langle r^2(t)\rangle \sim r_0^2\,e^{2\lambda t}, \qquad \lambda\sim\eta^{1/3},
\label{eq:exp}
\end{equation}
not algebraically~\cite{BoffettaSokolov2002,BoffettaEcke2012}. We stress a subtlety that Ref.~\cite{Chen2006} does not address: its own cited experimental support for Richardson-type dispersion in two-dimensional turbulence, Ref.~\cite{JullienParetTabeling1999}, probes the \emph{inverse} energy-cascade range (large scales, where 2D dynamics is transiently local and approximately 3D-like), not the forward enstrophy range relevant to small-scale dissipation and to the regularity argument developed below. The two ranges of 2D turbulence obey different dispersion laws, and only one of them is the analogue of the 3D inertial range that motivates $s=1/3$.

No finite-index stable process reproduces Eq.~(\ref{eq:exp}): the moments of an $\alpha$-stable process grow as a power of $t$ for every $\alpha\in(0,2]$, and exponential growth is not a member of that family at any finite $\alpha$; it is approached, formally and singularly, only in the degenerate limit $\alpha\to0$, which lies outside the normalizable stable laws on which the entire fractional-diffusion construction rests~\cite{Feller1971}. Chen's generative procedure --- solve for the stable index that reproduces the measured pair-dispersion law --- therefore does not yield a different answer in the enstrophy range: it yields \emph{no answer at all}. The method does not fail quietly; it has no fixed point to fail at.

\subsection{The calibrated equation does not know this}

Equation~(\ref{eq:chen}), once its coefficients $\nu,\gamma$ and its order $s=1/3$ are fixed numerically from 3D phenomenology, is nonetheless a well-defined operator in any spatial dimension $n$: the fractional Laplacian kernel is normalizable for all $n$ and $s\in(0,1)$. Nothing in the written equation switches off automatically when $n=2$, nor does it re-derive new coefficients from 2D statistics --- indeed, as just shown, no such re-derivation within Chen's own framework is possible. Solved literally in a two-dimensional domain with its 3D-calibrated coefficients, Eq.~(\ref{eq:chen}) would retain a permanent, non-local anomalous-dissipation channel at every $k>0$, with no phenomenological justification and in tension with the classical global-regularity theory of two-dimensional flow: conservation of enstrophy in the inviscid limit is precisely what guarantees that the velocity field remains in $H^1(\Omega)$ for all time, without recourse to any anomalous dissipation mechanism~\cite{Ladyzhenskaya1959,Yudovich1963,MajdaBertozzi2002}. An operator carried over unchanged from a regime where such a mechanism is required, into a regime where regularity is already guaranteed without it, is not a prediction; it is an artifact of not re-deriving the model.

\subsection{AFNS-IR's limit is not bolted on}

Section~V.B of Ref.~\cite{Lopez2026} shows $\mathrm{Re}_c\to\infty$ as $n\to2$ directly from the critical-Reynolds-number expression [Eq.~(33) of Ref.~\cite{Lopez2026}] in the limit $s_{\min}\to1$. This is not a separate calculation grafted onto the 3D result: the same variational free-energy structure that produces $s_{\min}=1/3$ under the dissipative-anomaly constraint in 3D relaxes, under the enstrophy constraint, to the trivial fixed point $s_{\min}=1$ in 2D. The order parameter itself is what deforms; asking it to relax to the local operator in the absence of vortex stretching requires no new physics, only the same free energy evaluated at $n=2$. By contrast, obtaining the analogous limit for Chen's construction would require an entirely new derivation from 2D pair-dispersion statistics --- which Sec.~IV.A shows does not exist within the stable-process ansatz the model is built on.

It is worth noting, in fairness to Ref.~\cite{Chen2006}, that its own internal bookkeeping already points in this direction without ever stating it. Chen's relation between the spectral exponent $\beta$ and the structure-function exponent $q$, $\beta=q+1$ [Sec.~2 of Ref.~\cite{Chen2006}], gives $\beta=3$ at $q=2$ --- his own smooth, non-Gaussian-velocity limit [Eq.~(12) of Ref.~\cite{Chen2006}] --- which coincides numerically with the Kraichnan--Batchelor--Leith enstrophy exponent. In Chen's own scheme, this limit is exactly where the governing equation reduces to the ordinary Laplacian [Eq.~(11) of Ref.~\cite{Chen2006}], not the fractional one. The qualitative direction is present, buried, in the 2006 paper; what is missing is the dimensional argument, and the re-derivation, that would make it operative in two dimensions.

\subsection{Assessment}

Applied to the enstrophy-conserving limit, the two frameworks are not merely quantitatively different --- they fail or succeed by different logical routes. AFNS-IR reproduces $\mathrm{Re}_c\to\infty$ as the smooth continuation of the same free-energy argument that fixes $s_{\min}=1/3$ in 3D. Chen's equation, if solved literally with its 3D-calibrated coefficients, reproduces nothing: it retains an unjustified non-local term in a regime where its own generative logic --- honestly extended --- yields no fractional order at all. This is a stronger statement than a failure to match a number: it is a demonstration that the crossover architecture of Eq.~(\ref{eq:chen}), unlike the deformation architecture of Eq.~(\ref{eq:afns}), does not carry its own justification across the dimensional boundary where the physics it was built on ceases to operate.

\section{Discussion}

The two-dimensional limit examined in Sec.~IV is, we would argue, a sharper discriminant than any measurement of the spectral crossover in three dimensions, precisely because dimensionality removes vortex stretching by kinematic necessity rather than by dynamical assumption. Neither model can evade the consequence: AFNS-IR's order parameter relaxes to the classical operator through the same variational logic that fixed it away from 1 in 3D, while Chen's generative mechanism, honestly extended, has no solution to relax to. Chen's construction remains, in its own domain of validity, a legitimate and useful low-order closure for wall-bounded 3D turbulence, and its citation record over the intervening two decades~\cite{Chen2006,Song2021,Epps2018,Suzuki2022} confirms its practical value independent of this question. What Sec.~IV shows is narrower and, we would argue, sharper: the two constructions are not interchangeable parametrizations of the same physics away from the regime both were tuned on, and the dimensional boundary at $n=2$ is where that difference becomes decidable without any new measurement, numerical or experimental.

This is precisely analogous to the distinction, in critical phenomena, between a crossover between two fixed universality classes and a genuine renormalization-group flow to a new fixed point~\cite{LandauLifshitz}: the two are often numerically indistinguishable over a finite window, and only the behavior at the boundary of the flow's domain --- here, the dimensional boundary rather than a scale boundary --- tells them apart. We note, finally, that both routes converge on $1/3$ in three dimensions while starting from unrelated formal machinery: a Lévy-flight stochastic argument in one case, a variational free-energy functional tied to the dissipative anomaly in the other. This convergence is evidence that $s=1/3$ is very likely a genuine attractor for \emph{any} fractional-Laplacian representation of three-dimensional inertial-range dissipation --- but it is evidence \emph{internal} to the fractional-Laplacian paradigm, not an external confirmation of that paradigm against the alternative, non-fractional closures with which it competes~\cite{Smagorinsky1963}.

A genuinely open question, which we leave for future work, is whether this dichotomy leaves any residual signature in real quasi-two-dimensional laboratory flows --- soap-film turbulence, or rapidly rotating and strongly stratified layers, in which vortex stretching is suppressed, if imperfectly, by physical rather than mathematical means. We deliberately do not propose a direct numerical simulation of the full Navier--Stokes equations as an arbiter of this dichotomy: the exact spectral energy balance in a statistically stationary flow is fixed by conservation alone, independently of any closure hypothesis, and a comparison built on it risks mistaking a bookkeeping identity for a test of physics. Whether a laboratory realization of the enstrophy-conserving limit shows any trace of anomalous, non-local dissipation, or none at all, is an experimental question we leave explicitly open.

\begin{acknowledgments}
The author thanks constructive discussions that motivated the comparison presented here.
\end{acknowledgments}

\end{document}